**Радиевый институт им. В. Г. Хлопина**

Победоносцев Л. А.

# ОТДЕЛЬНЫЕ ВОПРОСЫ КЛАССИЧЕСКОЙ ЭЛЕКТРОДИНАМИКИ

Санкт-Петербург – 2010

*В данной работе делается попытка с классических позиций объяснить ряд положений, лежащих в основе квантовой механики. При этом постоянная Планка появляется как следствие требования стабильности нуклона. Протон может быть представлен как вращающийся диск, плотность массы которого плавно убывает от центра к "ободу", на котором сосредоточен его заряд. В этом случае расчётный магнитный момент протона совпадает с его экспериментальным значением. Классическим путем получена формула Бальмера; оказывается, что радиусы протона и электронной орбиты в атоме водорода взаимосвязаны.*

*In the present work an attempt to explain with the classical stands a number of statements of the quantum mechanics has been done. At this the Plank constant appears as consequence of demand of nucleon stability. Proton can be imagined as a rotating disk which mass density is smoothly decreasing from centre to edge where its charge is concentrated. In this case the calculated magnetic moment of proton is equal to its experimental value. The Balmer formula has been derived with classical approach. It turns out that radii of electron orbit and proton in hydrogen atom are interconnected.*

# Введение

Напряженность электрического поля внутри тела определяется рядом факторов: распределением заряда в теле, формой и свойствами вещества тела и т. д. В частности, для шара радиуса R, имеющего равномерное по объему распределение заряда Q и массы M, согласно [1] при $r \leq R$:

$$E = \frac{Qr}{4\pi R^3}. \qquad (1)$$

Поместим в этот шар некий объект массой m и зарядом q, знаки зарядов Q и q пусть будут разные. Природа объекта (qm) не конкретизируется: это может быть нечто, принудительно внесённое в шар, но может быть и флюктуацией вещества и заряда шара, возникшее вследствие столкновения или под воздействием внешних полей. Нетрудно заметить, что уравнение движения объекта (qm) внутри шара (MQ) совпадает с уравнением обычного осциллятора с частотой (циклической) колебаний:

$$\omega = \frac{1}{2R}\sqrt{\frac{Q}{\pi R}\left(\frac{q}{m}\right)}. \qquad (2)$$

При таком рассмотрении шар и колеблющийся в нём объект являются единой системой, связанной электростатическими силами. Однако колебания объекта (qm) происходят не в вакууме, а в веществе шара, и в какой-то форме они должны передаваться шару. Но тело, вещество которого находится в состоянии колебаний или других форм внутреннего движения, не может быть стабильным.

В рассматриваемом шаре электростатическая сила, ответственная за колебания объекта (qm), всегда направлена к центру шара. Поэтому она в случае вращения шара частично или полностью может быть компенсирована центробежными силами, которые будут направлены в противоположную сторону. Так, в диаметральной плоскости шара, перпендикулярной оси вращения, проходящей через центр шара, полная компенсация сил наступает при угловой скорости вращения шара, совпадающей с выражением (2). При большей скорости вращения шара объект (qm) будет «вытолкнут» из шара центробежными силами, т. е. такой объект не может находиться внутри шара. Если имеются два объекта, причем $(\frac{q}{m})_1 \geq (\frac{q}{m})_2$, то в случае вращения шара с частотой $\omega_1$, соответствующей первому объекту, второй объект заведомо не может находиться в шаре.

Таким образом, если в Природе существует максимально возможное соотношение $(\frac{q}{m})_{max}$ и шар вращается с частотой $\omega_{max}$, то «существование» в нём каких-то объектов (qm), флюктуаций и т. д. делается невозможным. Такая частота вращения шара становится гарантом его внутренней стабильности. Из всех известных нам объектов наибольшим значением $(\frac{q}{m})$ обладает электрон, поэтому выдвигается гипотеза, что в случае электрона реализуется $(\frac{q}{m})_{max}$, т. е.

$$(\frac{q}{m})_{max} = (\frac{q}{m})_e = 5{,}275 \cdot 10^{17} \text{ CGSE}. \qquad (3)$$

Для симметричного вращающегося тела интегралом движения является момент количества движения:

$$I = J\omega. \qquad (4)$$

Здесь: J – момент инерции вращающегося тела. Для шара $J = \frac{2}{5}MR^2$, где M – масса шара. Учитывая (2) и (3), имеем:

$$I = \frac{1}{5}M\sqrt{\frac{QR}{\pi}(\frac{q}{m})_{max}}. \qquad (5)$$

## 1 Ядерные масштабы

Перенесём всё вышесказанное на протон: M = 1,673·10⁻²⁴ гр. При определении радиуса протона возникают нюансы: само понятие «радиус

протона» не является достаточно определённым, в литературе указываются несколько различные величины [2], но в любом случае он заключен в диапазоне $(1,2 – 1,8) \cdot 10^{-13}$ см.

Подставляем в выражение (5) значение $R=1,226 \cdot 10^{-13}$ см и указанные значения M и $(\frac{q}{m})_{max}$. Получаем: $I = 1,05 \cdot 10^{-27}$ эрг.сек – постоянная Планка h. Этот результат означает, что постоянная Планка не может рассматриваться как независимая физическая константа. Несмотря на то, что в работе имеются неопределённости и упрощения, представляется невозможным, что четыре независимых величины: Q, M, R, $(\frac{q}{m})_{max}$ случайно могут образовать комбинацию, в результате которой получается правильное значение другой величины.

Следовательно, в этом имеется физический смысл.

## 2. Реальный протон

Однако у протона спин равен $\frac{h}{2}$, а не h (h – постоянная Планка). Это указывает на то, что по «форме» протон не является шаром с равномерным по объёму распределением массы и заряда. Первоначальная модель нуждается в изменениях. Действительно, любое тело при вращении должно изменить свою форму:

Центробежные силы по отношению к его веществу могут рассматриваться как силы «растягивания» тела в направлениях, перпендикулярных оси вращения. Особенно это относится к телам, вещество которых является легкодеформируемым, – при вращении они приобретают «дискообразную» форму, симметричную относительно оси вращения. Но даже Галактики в результате вращения становятся «дискообразными» [4].

Симметричное тело вращения, имеющее ту же массу M и вращающееся с той же частотой $\omega$, что и ранее рассмотренный шар, но имеющее в два раза меньший спин, должно обладать в два раза меньшим моментом инерции (4). Это возможно, если плотность вещества тела убывает с возрастанием радиуса вращения и если его форма отлична от шара. Физически такое уменьшение плотности понятно: «растягивающие» силы, являясь по существу центробежными, возрастают с увеличением радиуса вращения; этим же объясняется и «дискообразная» форма. Далее используем самый простой вид зависимости плотности вещества от радиуса вращения, – линейный:

$$\rho = \rho_0 (1 - \frac{1}{R} r). \qquad (6)$$

Здесь: r – радиус вращения элемента тела; R – радиус тела (максимальное значение r) $\rho_0$ = const внутри тела; $\rho_0 = 0$ вне тела.

Представим «дискообразную» форму объекта в самом простом виде: два сложенных основаниями правильных конуса; таким образом, в сечении получается правильный ромб.

Рассчитаем массу и момент инерции такой «конструкции» в соответствии с Рис. 1. На этом рисунке ось вращения совпадает с координатой x; радиус вращения r перпендикулярен оси x. «Толщина» диска – $2\ell$; $2\alpha$ – угол при вершине диска; $tg\alpha = \frac{R}{l}$.

Элементу площади в плоскости рисунка dS = dxdr в пространстве соответствует элемент объёма dV = $2\pi$rdS и, соответственно, элемент массы dM = $\rho$dV = $2\pi\rho_0(1 - \frac{1}{R}r)$rdrdx. Таким образом, масса «полудиска» будет:

$$M = 2\pi\rho_0 \int_0^l dx \int_0^{\frac{R}{l}x} r(1 - \frac{r}{R}) dr = \frac{1}{6}\pi\rho_0 \ell R^2. \qquad (7)$$

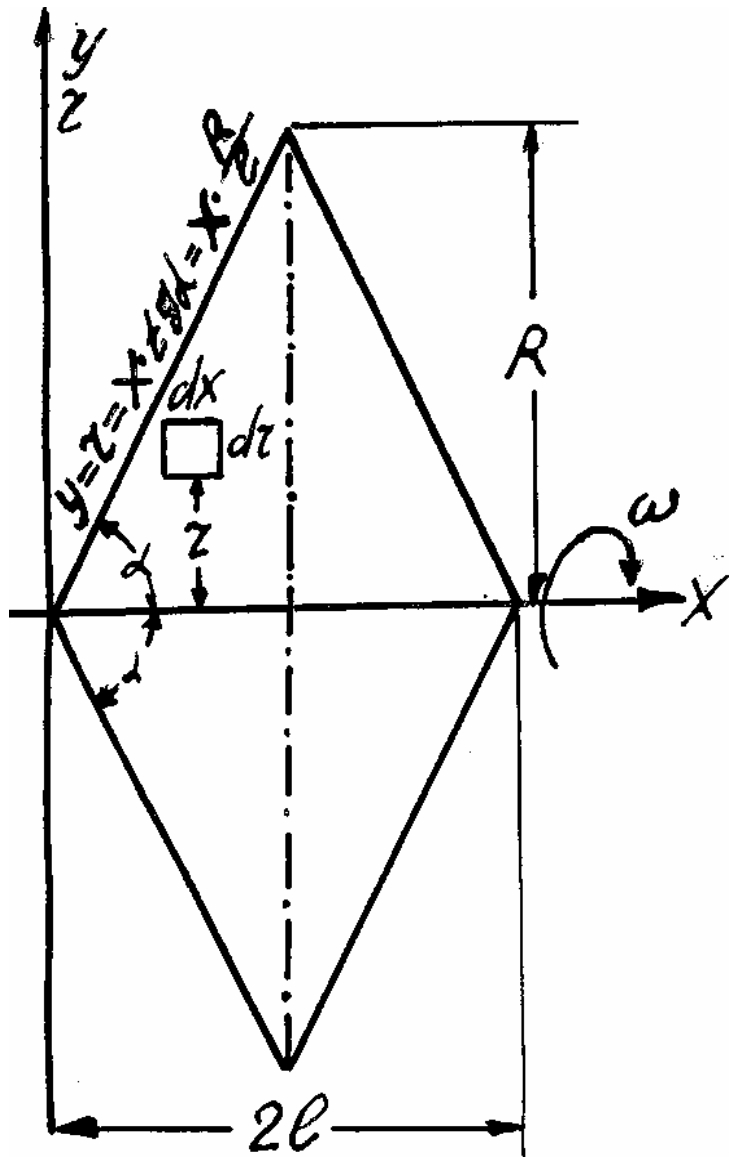

Рис. 1. Сечение диска.

R – радиус диска; r – радиус вращения элемента диска; 2$\ell$ – "толщина" диска; 2$\alpha$ – угол при вершине диска.

Так же рассчитываем момент инерции: $dJ = r^2\, dM = 2\pi\rho_0\, r^3\, (1-\frac{1}{R}r)\, dr\, dx$.

Интегрируя это выражение в тех же пределах, имеем:

$$J = \frac{1}{30}\pi\, \rho_0 \ell\, R^4. \tag{8}$$

Из (7) $\pi\rho_0 \ell = 6MR^{-2}$. Подставляя это в (8), получаем:

$$J = \frac{1}{5}MR^2. \qquad (9)$$

Момент инерции такой «дискообразной» фигуры составляет $\frac{1}{2}$ от момента инерции шара ($J = \frac{2}{5}MR^2$), что при прочих равных условиях, согласно (4), обеспечивает спин протона $\frac{1}{2}h$ (h – постоянная Планка). Рассмотрение выполнено для «полудиска», но ввиду симметрии фигуры для целого диска будет то же выражение (это равносильно умножению на 2 левой и правой части). Интересной особенностью рассмотренной «конструкции» является то, что конечный результат не зависит от «толщины» диска $2\ell$.

### 3. Магнитный момент протона

С распределением массы протона M связан его момент вращения; но у протона имеется также заряд Q = 4,803·10⁻¹⁰ CGSE.

Электрический заряд не существует без массы, а масса в данном рассмотрении вращается с частотой $\omega$ (2); поэтому и заряд, вероятней всего, вращается с такой же частотой. Вращающийся заряд создаёт электрический ток, а с ним уже связан магнитный момент, являющийся экспериментально определяемой величиной. Поэтому сопоставление действительной (экспериментально определённой) величины магнитного момента с его расчётным значением может служить критерием верности расчётных предпосылок.

«Ядерный магнетон», рассчитанный в соответствии с канонами квантовой механики, $\mu_0$ = 5,05·10⁻²⁴ эрг/Гаусс. В действительности у протона магнитный момент $\mu_p = 2,79\mu_0 = 1,41\cdot 10^{-23}$ эрг/Гаусс.

В прошлом утверждалось, что магнитный момент протона имеет другое, неэлектромагнитное происхождение; однако не пояснялось, какие «другие» процессы имеются в виду.

Далее рассчитаем радиус круговой орбиты r, при движении по которой с частотой $\omega$ (2) заряд протона Q создаёт действительный магнитный момент протона $\mu_p$. В этой задаче безразлично, заряд сосредоточен в одной точке или как-то «размазан» по орбите; также r может быть не чётким радиусом в геометрическом понимании, а «эффективным радиусом» некоторого распределения. Исходим из общего выражения для магнитного момента: $\mu = \frac{j}{c} S$ [2], где S – площадь, охватываемая контуром с током j; c – скорость света.

В рассматриваемом случае $S = \pi r^2$; $j = Q\gamma$; $\gamma$ – частота вращения заряда по орбите радиуса r ($\gamma = \frac{\omega}{2\pi}$); $j = \frac{Q\omega}{2\pi}$. Сделав подстановки, имеем:

$$\mu = \frac{Q\omega}{2c} r^2, \text{ или } r = \sqrt{\frac{2\mu c}{Q\omega}}. \qquad (10)$$

Подставляя в (10) значение $\omega$, согласно (2), окончательно имеем:

$$r = 2(\mu c)^{1/2} \left[ \frac{1}{\pi} \left(\frac{Q}{R}\right)^3 \left(\frac{q}{m}\right)_{max} \right]^{-1/4}. \qquad (11)$$

Используем следующие значения: $\mu = \mu_p = 1{,}41 \cdot 10^{-23}$ эрг/Гаусс – действительный магнитный момент протона; $c = 3 \cdot 10^{10}$ см/с – скорость света; $Q = 4{,}803 \cdot 10^{-10}$ CGSE – заряд протона (элементарный заряд); $R = 1{,}226 \cdot 10^{-13}$ см – значение радиуса протона при определении спина; $\left(\frac{q}{m}\right)_{max} = 5{,}275 \cdot 10^{17}$ CGSE, – как и ранее.

Подставляя эти значения в (11), имеем: $r = 1{,}2975 \cdot 10^{-13}$ см. Это значение r всего на 5,5% больше R. Такое различие вполне можно объяснить делаемыми в работе упрощениями. Таким образом, форму протона можно представить в виде диска с плавно убывающей плотностью массы от центра (ось вращения) к "ободу", на котором сосредоточен заряд, создающий круговой ток в «плоскости протона».

Ядра же состоят из протонов и нейтронов. И чтобы получить для них распределения массы и заряда, необходимо выполнить усреднения по всем протонам и нейтронам, входящим в ядро. Тогда, какими бы ни были распределения массы и заряда в нуклонах, при их достаточном количестве в ядре в результате усреднения мы получим равномерные распределения. Что и подтверждается экспериментально [2].

### 4. Внешний электрон – атом водорода

Далее рассмотрим движение внешнего электрона в поле такого "дискообразного" протона. Это рассмотрение во многом сходно с теми, которые приводятся в многочисленных изданиях при рассмотрении "кеплеровой задачи". Но мы попытаемся ответить на вопрос, как на системе "дискообразный" протон – внешний электрон (т. е. атом водорода) отражаются магнитные поля, создаваемые их движениями.

Конечно, обычное кулоновское взаимодействие будет преобладающим. Орбиты электрона в атоме водорода располагаются в одной плоскости и имеют форму эллипса, в одном из фокусов которого располагается протон. В частности, орбита электрона может быть окружностью радиуса r; для упрощения последующих расчётов мы будем рассматривать этот случай. Из условия равновесия кулоновских и центробежных сил, действующих на электрон, определяется скорость его движения на орбите: $V = \frac{e}{\sqrt{rm}}$.

Здесь: e – заряд электрона, m – его масса. Соответственно этому частота вращения электрона на орбите:

$$\gamma = \frac{V}{2\pi r}; \text{ циклическая частота } \omega = \frac{e}{r\sqrt{mr}}. \quad (12)$$

Вращающийся на орбите электрон создаёт круговой ток: $j_e = e\gamma = \frac{e^2}{2\pi r\sqrt{mr}}$.

В свою очередь ток создаёт магнитное поле, напряженность которого в центре электронной орбиты, где расположен протон [4]

$$H_e = \frac{2\pi j_e}{cr} = \frac{e^2}{cr^2\sqrt{mr}}. \quad (13)$$

Направление магнитного поля $H_e$ в центре электронной орбиты перпендикулярно плоскости орбиты и совпадает с осью вращения.

Кроме кулоновской и центробежной сил на вращающийся электрон должна действовать также сила Лоренца, так как в области орбиты электрона протон создаёт магнитное поле напряженностью $H'_p$

$$F_l = \frac{e}{c}[VH'_p].$$

Сила Лоренца компенсируется кариолисовой силой [5] $F_k$, что проявляется во вращении плоскости орбиты электрона (оси вращения орбиты электрона) с частотой $\Omega_e$ вокруг направления $H'_p$ – ларморова прецессия: $F_k = 2m[V\Omega_e]$. Из равенства $F_l = F_k$ имеем:

$$\Omega_e = \frac{e}{2cm}H'_p. \quad (14)$$

Для определения $\Omega_e$ необходимо установить $H'_p$, которое создаётся вращением заряда "дискообразного" протона с частотой $\omega$ (2).

В нашем рассмотрении оси вращений орбиты электрона и заряда "дискообразного" протона почти параллельны (или антипараллельны), плоскости же их вращений почти совпадают. Но если протон находится в центре электронной орбиты, то электрон находится вне токовой орбиты "дискообразного" протона – на расстоянии r от его центра. Эта "география" напоминает одно кольцо внутри другого в одной плоскости (Рис 2). На этом рисунке не изображён угол между осями вращения, без которого само понятие "прецессия" теряет смысл. Но этот угол никак не вошел бы в конечные результаты, и он, очевидно, мал.

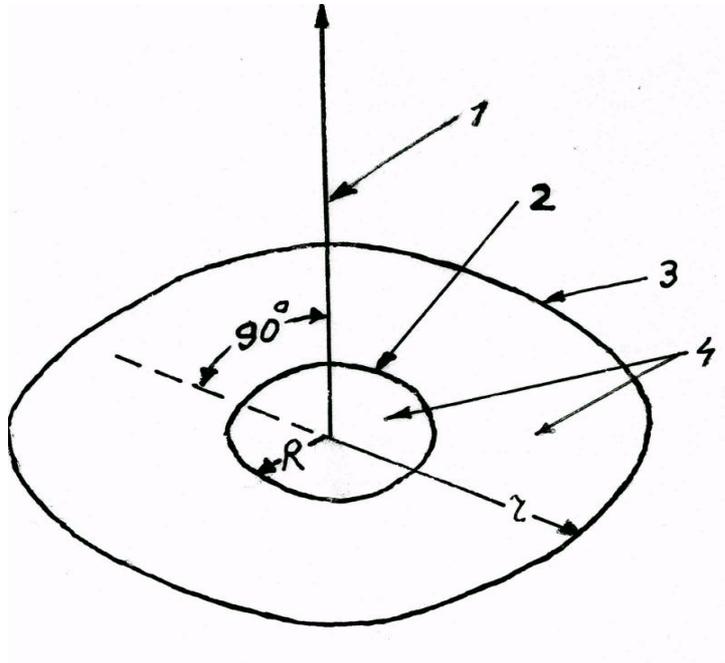

Рис. 2.

1 – ось вращения; 2 – орбита заряда "дискообразного" протона; 3 – орбита внешнего электрона; 4 – плоскости вращения.

Для тока $j_p$, протекающего по окружности радиуса R (в нашем случае это радиус протона), напряженность магнитного поля на оси окружности на расстоянии d от её центра при d>>R даётся выражением: $H = \dfrac{2\pi j_p R^2}{cd^3}$ [4].

Несложно определить и напряженность магнитного поля в плоскости окружности с током [6], в данном случае:

$$H'_p = \dfrac{\pi j_p R^2}{cd^3}. \qquad (15)$$

В нашем случае: $j_p = \dfrac{e\omega}{2\pi}$; $\omega$ даётся выражением (2); d = r – радиус орбиты электрона. Сделав соответствующие подстановки в (14), получаем:

$$\Omega_e = \dfrac{e^3\sqrt{R}}{8c^2 r^3 \sqrt{\pi m^3}}. \qquad (16)$$

Сходные рассуждения могут быть применены и к "дискообразному" протону. Поскольку заведомо r>>R, можно считать, что протон находится в центре электронной орбиты, в котором напряженность магнитного поля от электронной орбиты даётся выражением (13), сила Лоренца, воздействующая на "дискообразный" протон:

$$F_{lp} = \frac{e}{c}[V_P H_e] = \frac{e^3 V_P}{c^2 r^2 \sqrt{mr}}. \qquad (17)$$

Здесь $V_p$ – скорость на ободе "дискообразного" протона, на котором сосредоточен его заряд.

Однако кариолисову силу в случае протона $F_{kp}$ мы не можем прямо записать в таком же виде, как для электрона.

Для электрона его заряд и масса принимаются сосредоточенными в одной точке; в случае же протона его масса M рассредоточена по объёму "дискообразной" формы. Поэтому необходима некоторая коррекция. Однако для элемента массы протона dM элемент кариолисовой силы $dF_{kp}$ может быть записан как для электрона

$$dF_{kp} = 2dM V_1 \Omega_p. \qquad (18)$$

Здесь: $V_1$ – скорость, соответствующая элементу массы dM, $\Omega_p$ – скорость угловой прецессии "дискообразного" протона. Очевидно:

$$V_1 = \frac{V_p}{R} r_1, \qquad (19)$$

где $r_1$ – расстояние от оси вращения "дискообразного" протона до элемента его массы dM.

Как и в случае подсчета массы, спина и магнитного момента "дискообразного" протона: $dM = 2\pi\rho_0 (1 - \frac{r_1}{R}) r_1 \, dr_1 \, dx$. Подставляя эти значения, а также (19) в (18), получаем:

$$dF_{kp} = \frac{4\pi\rho_0 \Omega_p}{R} V_p (1 - \frac{r_1}{R}) r_1^2 \, dr_1 \, dx. \qquad (20)$$

Интегрируя это выражение в тех же пределах, что и ранее, имеем:

$$F_{kp} = \frac{4}{5} V_p \, M \, \Omega_p. \qquad (21)$$

Теперь мы можем, как и в случае электрона, для "дискообразного" протона записать $F_{lp} = F_{kp}$ и в результате получаем:

$$\Omega_p = \frac{5e^3}{4c^2 M r^2 \sqrt{mr}}. \qquad (22)$$

Здесь: r – радиус электронной орбиты; m – масса электрона; M – масса протона; e – элементарный заряд; c – скорость света.

Таким образом, у нас получилась "взаимовращающаяся" система: плоскость электронной орбиты прецессирует в соответствии с магнитным полем, создаваемым "дискообразным" протоном (16). А протон, в свою очередь, прецессирует в соответствии с магнитным полем, создаваемым орбитальным электроном (22).

По своей величине скорости угловых прецессий $\Omega_e$ и $\Omega_p$ существенно меньше скоростей вращения электрона на орбите (12) и "дискообразного" протона вокруг своей оси (2). Но даже небольшие воздействия в случае их периодичности могут приводить к существенным результатам для системы в целом. Математически строгое и последовательное описание такой системы представляется весьма сложным. Но если в целом такая динамическая система является стабильной, то её состояния должны (или могут) периодически повторяться. В данном случае это выглядит следующим образом: если "дискообразный" протон вследствие прецессии сделает полный оборот, то он займёт первоначальное положение. Но чтобы вся система "дискообразный" протон – внешний электрон заняла первоначальное положение, необходимо, чтобы плоскость электронной орбиты одновременно заняла бы тоже первоначальное положение. Это возможно лишь при определённом соотношении между частотами прецессии, а именно: $\Omega_p = N\Omega_e$, где N – целое число. (Более общий случай соотношения между частотами прецессии $N_p \Omega_p = N_e \Omega_e$, где $N_p$ и $N_e$ являются целыми числами, в данной работе не рассматривается). Мы ограничим данное рассмотрение четными значениями N, т. е. N = 2n. Формально – математически – это допустимо. Физически это удвоение, возможно, связано с тем, что фактически имеются две зеркальные комбинации: с параллельной и антипараллельной ориентацией орбитального момента электрона и спина протона. Мы же в данной работе эти два случая принимаем за один вариант. Имеем:

$$\Omega_p = 2n\Omega_e; \quad n = 1; 2; 3; 4 \text{ и т. д.} \qquad (23)$$

Подставляем в (23) $\Omega_e$ согласно (16) и $\Omega_p$ согласно (22). После простых и очевидных преобразований получаем:

$$r = \frac{R}{25\pi}\left(\frac{M}{m}\right)^2 n^2. \qquad (24)$$

Получилось простое соотношение. Оно связывает ядерные масштабы – радиус протона R – с атомными масштабами – радиусом электронной орбиты r; это своеобразный "мост" между ними. Отношение массы протона M к массе электрона m хорошо известная величина – $\left(\frac{M}{m}\right) = 1{,}83 \cdot 10^3$ – и ранее приведенное выражение можно записать в виде:

$$r = 4{,}27 \cdot 10^4 \, R \, n^2. \qquad (25)$$

Если в этом выражении при n = 1 использовать R = 1,2388·10⁻¹³ см, что хорошо согласуется с ранее используемыми значениями R, то получим r = 5,29·10⁻⁹ см. Это значение r совпадает с радиусом первой боровской орбиты, соответствующей атому водорода в основном состоянии, – величиной давно и хорошо известной.

Далее последовательно получаем для атома водорода:

1. Скорость движения электрона на орбите

$$V = \frac{e}{\sqrt{rm}} = \frac{2{,}189}{n} 10^8 \text{ см/сек.}$$

2. Механический момент орбитального вращения электрона

$$P = mVr = 1{,}05 \cdot 10^{-27} \text{ n эрг.сек},$$

где $1{,}05 \cdot 10^{-27}$ эрг.сек. – постоянная Планка $h$. У нас опять получилась постоянная Планка, но уже другим путем.

Таким образом, все вышеизложенное становится самосогласованной системой (выражение P = hn постулировалось в первоначальной теории Бора).

3. Кинетическая энергия орбитального электрона

$$E = \frac{mV^2}{r} = \frac{2{,}18}{n^2} 10^{-11} \text{ эрг} = \frac{13{,}6}{n^2} \text{ эВ.}$$

4. Разность энергий между состояниями электрона с разными значениями n

$$\Delta E = 13{,}6 \left(\frac{1}{n_1^2} - \frac{1}{n_2^2}\right) \text{ эВ.}$$

Это одна из форм записи соотношения Бальмера.

Для соответствия расчётных значений их экспериментальным данным в работе использовались следующие значения радиуса протона R:

1. Для механического момента вращения (спина) абстрактного шара и "дискообразного" протона:

    $R_1 = 1{,}226 \cdot 10^{-13}$ см.

2. Для магнитного момента "дискообразного" протона:

    $R_2 = 1{,}2975 \cdot 10^{-13}$ см.

3. Для радиуса электронной орбиты атома водорода:

    $R_3 = 1{,}2388 \cdot 10^{-13}$ см.

    Усредняя эти значения, имеем: $R_{ср} = (1{,}25 \pm 0{,}03) \cdot 10^{-13}$ см.

## Заключение

В данной работе все построения и расчеты выполнены классическими методами. На основании полученных результатов можно сделать следующие выводы:

1. Вопреки распространенному мнению о том, что некоторые экспериментальные результаты не могут быть объяснены с классических позиций, – например, формула Бальмера, – в действительности имеют классическое объяснение.

2. Обнаружена зависимость радиуса электронной орбиты в атоме водорода от радиуса протона.

3. Достигается согласие между расчетным и экспериментальным значениями магнитного момента протона.

## Литература

1. А. С. Кингсен, Г. Р. Локшин, О. А. Ольфов. Основы физики. Курс общей физики. М.: Физматлит., 2001, т. 1, с. 174.

2. Л. Хюльтен, М. Сугавара. Строение атомного ядра / Под ред. А. С. Давыдова. М.: Издат. иностранной литературы, 1959, с. 63 – 69, 272.

3. Ч. Киттель, У. Найт, М. Рудерман. "Берклиевский курс физики". Т. 1, "Механика". М.: Наука, 1971, с. 212, рис. 6, 24.

4. И. Е. Тамм. Основы теории электричества. М.: Гос. издат.-во технико-теорет. литературы, 1956, с. 207.

5. Э. В. Шпольский. Атомная физика. Т. 1. М.: Наука, с. 244.

6. Л. А. Победоносцев. Постоянная Планка – атом водорода. Препринт Радиевого института им. В. Г. Хлопина РИ-261. Приложение 1. Санкт-Петербург, 2004.